\documentclass[conf]{new-aiaa}
\usepackage[utf8]{inputenc}
\usepackage{textcomp}
\usepackage{graphicx}
\usepackage{amsmath}
\usepackage[version=4]{mhchem}
\usepackage{siunitx}
\usepackage{longtable,tabularx}
\usepackage{float}
\usepackage{multirow}
\usepackage{comment}
\usepackage{mathtools}
\usepackage{autobreak}

\newtheorem{theorem}{Theorem}
\newtheorem{lemma}[theorem]{Lemma}

\title{Pulse Width Modulation Method Applied to Nonlinear Model Predictive Control on an Under-actuated Small Satellite}

\author{Kota Kondo \footnote{Student, Department of Mechanical and Aerospace Engineering, AIAA Student Member.} and Yasuhiro Yoshimura\footnote{Assistant Professor, Department of Aeronautics and Astronautics.}}
\affil{Kyushu University, Fukuoka-shi, Fukuoka, 819-0395, Japan}
\author{Shuji Nagasaki\footnote{Assistant Professor, Department of Aeronautics and Astronautics.}}
\affil{Kyushu University, Fukuoka-shi, Fukuoka, 819-0395, Japan }
\author{Toshiya Hanada\footnote{Professor, Department of Aeronautics and Astronautics.}}
\affil{Kyushu University, Fukuoka-shi, Fukuoka, 819-0395, Japan }

\begin{document}

\maketitle

\begin{abstract}
Among various satellite actuators, magnetic torquers have been widely equipped for stabilization and attitude control of small satellites. Although magnetorquers are generally used with other actuators, such as momentum wheels, this paper explores a control method where only a magnetic actuation is available. We applied a nonlinear optimal control method, Nonlinear Model Predictive Control (NMPC), to small satellites, employing the generalized minimal residual (GMRES) method, which generates continuous control inputs. Onboard magnetic actuation systems often find it challenging to produce smooth magnetic moments as a control input; hence, we employ Pulse Width Modulation (PWM) method, which discretizes a control input and reduces the burden on actuators. In our case, the PWM approach discretizes control torques generated by the NMPC scheme. This study's main contributions are investigating the NMPC and the GMRES method applied to small spacecraft and presenting the PWM control system’s feasibility. 
\end{abstract}

\section*{Nomenclature}

{\renewcommand\arraystretch{1.0} 
\noindent\begin{longtable*}{@{}l @{\quad=\quad} l@{}}

$\boldsymbol{B}$, $\boldsymbol{B_0}$   & Earth's magnetic field vector in body and orbital frame, respectively\\ 
$H$ & Hamiltonian \\
$i$ & inclination \\
$J_\text{cost}$  & cost function\\
$\boldsymbol{J}$  & moment of inertia of satellite\\
$\boldsymbol{m}$ & magnetic dipole moment\\
$u_\text{max}$   & maximum control input of magnetic torquer \\
$N$ & discretized step number on prediction horizon\\
$\boldsymbol{q}$ & quaternion \\ 
$\boldsymbol{Q}, R$ & weight function \\
$\boldsymbol{Q_{\rm t}}$ & terminal cost \\
$r$ & distance from the center of Earth\\
$\boldsymbol{T}$  & control torque vector  \\
$T_s$   & final time on prediction horizon \\
$v$ & dummy input\\
$\lambda$ & Lagrange multiplier\\
$\boldsymbol{\omega}$ & angular velocity vector\\
$\omega_e$ & argument of perigee\\
$\theta$ & true anomaly\\
\multicolumn{2}{@{}l}{Subscripts}\\
$i$ & $i$-th time step on NMPC prediction horizon \\
$*$ & conditions on prediction horizon\\

\end{longtable*}}

\section{Introduction}

\lettrine{S}{tabilization} of angular velocities and attitude control is a crucial phase for satellite missions. There have been many approaches to detumbling and controlling satellite attitude. Various actuation systems, such as thrusters, momentum wheels, and control moment gyros~\cite{thrusters, momentum, momemt_gyro}, are being equipped with spacecraft. However, in the case of small satellites, due to restrictions on their size, weight, and budget, available actuators are significantly limited. Thus, magnetic torquers, which create magnetic dipole moment as a control input, interfering with an Earth's magnetic field, have been intensely studied. Magnetorquers are relatively small and inexpensive, and hence, they can save size and costs, which often restrict small satellite design. Nonetheless, magnetic actuators require dense magnetic flux, and their dependency on magnetic environments makes the control system under-actuated~\cite{magnetic_3, magnetic_4}. 

Severe restrictions are imposed on not only actuators but also on-board central processing units (CPUs). An extreme space environment - extensive solar radiation and significant temperature changes- limits CPUs' computational capability. However, feedback optimal control algorithms such as NMPC require a large-scale on-board calculation. Therefore, it is necessary to employ a method that reduces calculation cost and allows us to find control inputs faster. To overcome the problem, we apply the generalized minimal residual (GMRES) method~\cite{Ohtsuka_A_continuation/GMRES_method_for_fast_computation_of_nonlinear_receding_horizon_control}
, which is incorporated in the NMPC algorithm and helps find optimal solutions in a small amount of time.

However, the NMPC controller finds continuous and smooth optimal control inputs instead of discrete values, which causes an additional burden on control actuators. Therefore, our research examines the Pulse Width Modulation (PWM) method~\cite{Holtz_Pulsewidth_modulation_a_survey}, which discretizes the smooth inputs at every sampling time. Some satellites employ the PWM method~\cite{Massey_Continuous_Traditional_and_High-Order_Sliding_Modes_for_Satellite_Formation_Control}; however, to the best of our knowledge, there has not been a small satellite in which the control system finds optimal discrete inputs converted from continuous inputs optimized by the NMPC algorithm and the GMRES method. This discretization significantly lessens the complication of the actuation system. This advantage results in a more extended satellite operation, possible inexpensive actuators, and a robust control system.

The NMPC controller has some advantages when it comes to small satellite operations. Since it can consider control inputs in its cost function, it finds an optimal control, which decreases the use of magnetic moments as much as possible~\cite{Ohtsuka_2}. For small satellites, where available electric power is limited, this NMPC feature is significantly beneficial. Also, since the NMPC approach is a closed-loop control algorithm, it works well with the PWM method~\cite{Ohtsuka_hover}. The continuous control inputs the NMPC finds at the previous sample time are different from those of actual discrete inputs because the PWM coverts them into discrete values. However, the NMPC controller can adjust its control inputs in the next step and find the optimal inputs every sampling time.   

This paper explores the applicability of the GMRES method in the literature of magnetically actuated small spacecraft controlled by the NMPC and PWM control methodologies. The simulation section first showcases the detumbling approach, where a small satellite's rotation is attenuated. This phase is critical to move onto the next stage - attitude control, where we achieve the desired satellite attitude. In addition to simulation results, theoretical analyses on the controllability of such spacecraft have been made. As for controllability, we present that the magnetic torquer satellite system's control matrix is definite-positive on average, which concludes the system is controllable. 


\section{Rotational Kinematics and Dynamics}

Assume the origin of the body-fixed frame of a rigid-body satellite is located at the center of mass, and the principal axes of inertia are aligned with the frame axes. Then, the rotational motion of the spacecraft angular velocity vector $\boldsymbol{\omega}$ is described as the Euler's equations \cite{dynamics}:

\begin{equation}\label{dynamics_components}
    \boldsymbol{J}\boldsymbol{\dot{\omega}}=-\boldsymbol{\omega} \times \boldsymbol{J}\boldsymbol{\omega}+\boldsymbol{T}
\end{equation}
A three-axis magnetic torquer generates control torque $\boldsymbol{T}$ by interfering with the magnetic field~\cite{torque}:

\begin{eqnarray}\label{torque_in_matrix}
    \boldsymbol{T}
    &=&\boldsymbol{m} \times \boldsymbol{B}(t)\\
    &=&
    \begin{bmatrix}
        m_yB_z(t)-m_zB_y(t)\\
        m_zB_x(t)-m_xB_z(t)\\
        m_xB_z(t)-m_yB_x(t)\\
    \end{bmatrix}
\end{eqnarray}
Substituting Eq.~\eqref{torque_in_matrix} into Eq.~\eqref{dynamics_components}, we obtain

\begin{equation}\label{dynamics_combined}
    \begin{bmatrix}
        \dot\omega_{x}\\
        \dot\omega_{y}\\
        \dot\omega_{z}\\
    \end{bmatrix}
    =
    \begin{bmatrix}
        \frac{1}{J_x}\{(J_y-J_z)\omega_y\omega_z+m_yB_z(t)-m_zB_y(t)\}\\
        \frac{1}{J_y}\{(J_z-J_x)\omega_z\omega_x+m_zB_x(t)-m_xB_z(t)\}\\
        \frac{1}{J_z}\{(J_x-J_y)\omega_x\omega_y+m_xB_z(t)-m_yB_x(t)\}\\
    \end{bmatrix}
\end{equation}

Also, the rotational kinematics in the quaternion vector $\boldsymbol{q}$ is given in \cite{Magnetocoulombic}:

\begin{equation}
    \boldsymbol{\dot{q}}=E(\boldsymbol{q})\boldsymbol{\omega}
\end{equation}
where $\boldsymbol{q}=[q_1, q_2, q_3, q_4]^{T}$, whose norm is equal to 1, and $E(\boldsymbol{q})$ is given by

\begin{equation}
    E(\boldsymbol{q})
    =\frac{1}{2}
    \begin{bmatrix}
        q_4 & -q_3 & q_2\\
        q_3 & q_4 & -q_1\\
        -q_2 & q_1 & q_4\\
        -q_1 & -q_2 & -q_3\\
    \end{bmatrix}
\end{equation}


\section{Control Law}

\subsection{Controllability Analysis}

This section analyzes the controllability properties of the satellite angular velocity dynamics in the three-axis magnetic actuation. Note that we show the controllability by indicating that the control matrix $\psi_3$ is positive definite on average as indicated in~\cite{Magnetocoulombic}. 

\subsubsection{A three-axis magnetic actuator system}

Define a skew-symmetric matrix $S(\boldsymbol{\boldsymbol{\omega}})$  $\in$ $\mathbb{R}^{3\times 3}$ as
 \begin{equation}\label{skew_symmetric_matrix}
     S(\boldsymbol{\omega})=
     \begin{bmatrix}
        0 & \omega_z & -\omega_y\\
        -\omega_z & 0 & \omega_x\\
        \omega_y & -\omega_x & 0\\
     \end{bmatrix}
 \end{equation}
 and note that $S(\boldsymbol{\boldsymbol{\omega}})$ is not invertible.
 Then the vector of control torques in three-axis magnetic actuation is given by\cite{Fast_Terminal} 
 \begin{equation}\label{torque_using_skew_matrix}
    \boldsymbol{T}=\boldsymbol{m} \times \boldsymbol{B}
    =S(\boldsymbol{B}(t))\boldsymbol{m}
\end{equation}
Given the vector of the desired control inputs, $\boldsymbol{u} \in \mathbb{R}^3$, the magnetic moment $\boldsymbol{m}$ is computed  
using the pseudoinverse matrix~\cite{Pseudoinverse},
\begin{equation}
    \boldsymbol{m}=
    \frac{S(\boldsymbol{B}(t))^T}{||\boldsymbol{B}(t)||^2}\boldsymbol{u}
\end{equation}
Therefore, the actual vector of control torques is given by 
\begin{equation}
    \boldsymbol{T}=
    \frac{S(\boldsymbol{B}(t))S(\boldsymbol{B}(t))^T}{||\boldsymbol{B}(t)||^2}\boldsymbol{u}
\end{equation}
Letting $\boldsymbol{b}(t)=\frac{\boldsymbol{B}(t)}{||\boldsymbol{B}(t)||}$, it follows that
\begin{equation}\label{T_S_S_u_formulation}
\begin{array}{c}
    \boldsymbol{T}=S(\boldsymbol{b}(t))S(\boldsymbol{b}(t))^T\boldsymbol{u}=\psi_3(\boldsymbol{b}(t))\boldsymbol{u}\\
\end{array}
\end{equation}
where $\psi_3(\boldsymbol{b}(t))$ is given as
\begin{equation}
    \psi_3(\boldsymbol{b}(t))=
    \begin{bmatrix}
        b_y^2+b_z^2 & -b_y b_x & -b_x b_z \\
        -b_y b_x & b_x^2+b_z^2 & -b_y b_z \\
        -b_x b_z & -b_y b_z & b_y^2+b_x^2 \\
    \end{bmatrix}
\end{equation}    
Therefore, the dynamics of a satellite with a three-axis magnetic actuation is 
represented by
 \begin{equation}\label{dynamics_S_formulation}
    \boldsymbol{J} \boldsymbol{\dot\omega}=S(\boldsymbol{\omega}) \boldsymbol{J} \boldsymbol{\omega}+\psi_3(\boldsymbol{b}(t))\boldsymbol{u}
\end{equation}
Since the rank of $\psi_3$ is equal to $2$, the spacecraft dynamics are under-actuated at every instant of time. However, the system is controllable in case the control matrix $\psi_3$ is positive definite on  average~\cite{Magnetocoulombic}. By Lemma~\ref{controllability_lamma_1}, the average value of the magnetic control matrix $\Bar{\psi}_3$ is positive definite; therefore, the system given in Eq.~\eqref{dynamics_S_formulation} is controllable~\cite{Controllability_2}.

\begin{lemma}\label{controllability_lamma_1}

    Define $\boldsymbol{b_0}(t)$ as Earth's magnetic field unit vector with respect to an orbital frame. 
    Then, assuming $S(\boldsymbol{b_0}(t))\boldsymbol{\dot{b}}_0(t) \neq 0$ for all $t>0$,
    it follows for all $\tau>0$ that    
    \begin{equation}
        \hat{\psi}_{03}=\frac{1}{\tau}\int^\tau_0
        S(\boldsymbol{b_0}(t))S(\boldsymbol{b_0}(t))^T dt > 0
    \end{equation}
    Furthermore,
    \begin{equation}
        \Bar{\psi}_{03}=\lim_{\tau \rightarrow \infty }\frac{1}{\tau}\int^\tau_0
        S(\boldsymbol{b_0}(t))S(\boldsymbol{b_0}(t))^T dt > 0
    \end{equation}
    where $\Bar{\psi}_{03}$ is the average magnetic control matrix in the orbital frame. 

\end{lemma}{}

If $||\boldsymbol{\tau}_r||<\infty$, $\forall \tau >t_0$ where $0<t_0<\infty$ and $\boldsymbol{\tau}_r$ is the relative angular rate vector between body and orbital frame, then   

\begin{equation}
    \hat{\psi}_3=\frac{1}{\tau}\int^\tau_0
    S(\boldsymbol{b}(t))S(\boldsymbol{b}(t))^T dt > 0
\end{equation}

\begin{equation}
    \Bar{\psi}_3=\lim_{\tau \rightarrow \infty }\frac{1}{\tau}\int^\tau_0
    S(\boldsymbol{b}(t))S(\boldsymbol{b}(t))^T dt > 0
\end{equation}
The proof of Lemma \ref{controllability_lamma_1} is given in Ref.~\cite{Finite_time}. 


\subsection{NMPC Formulation}

This section describes the NMPC approach to detumbling a satellite with the nonlinear dynamics represented by Eq.~\eqref{dynamics_combined} based on the following receding horizon optimal control problem,
\begin{equation}\label{NMPC_formulation}
    \centering
    \begin{array}{rl}
    {\rm minimize} & J_\text{cost}
    =\frac{1}{2}(\boldsymbol{x}(t+T_s)-\boldsymbol{x}_{f})^{T}Q_{\rm t}(\boldsymbol{x}(t+T_s)-\boldsymbol{x}_{f})\\
    &+\int^{t+T}_t \frac{1}{2}\{(\boldsymbol{x}(\tau)-\boldsymbol{x}_{f})^{T}Q(\boldsymbol{x}(\tau)-\boldsymbol{x}_{f})+\boldsymbol{u}(\tau)^{T}R\boldsymbol{u}(\tau)-p_1v_1-p_2v_2-p_3v_3\}
    d\tau\\
    {\rm subject \ to} 
    & \boldsymbol{\dot{q}}=E(\boldsymbol{q})\boldsymbol{\omega}\\
    & \boldsymbol{J} \boldsymbol{\dot\omega}+
    \boldsymbol{\omega} \times \boldsymbol{J} \boldsymbol{\omega}=\boldsymbol{T}\\
    & \boldsymbol{T}=\boldsymbol{m} \times \boldsymbol{B}\\
    & m_x^2+v_x^2-u_\text{max}^2=0\\
    & m_y^2+v_y^2-u_\text{max}^2=0\\
    & m_z^2+v_z^2-u_\text{max}^2=0\\
    \end{array}
\end{equation}
where $\boldsymbol{x}=[q_1, q_2, q_3, q_4, \omega_x, \omega_y, \omega_z]^{T}$, $\boldsymbol{x}_f$ is the reference state vector, and $\boldsymbol{Q}$ and $R$ are positive-definite weight matrices, and $\boldsymbol{Q_{\rm t}}$ is terminal cost.
The auxiliary inputs, $v_x$, $v_y$, and $v_z$ are introduced following \cite{NMPC_formulation_1} to enforce the control constraints by recasting them as equality constraints in 
Eq.~\eqref{NMPC_formulation}. The negative sign preceding $p_x$, $p_y$, and $p_z$ in the cost function being minimized promotes keeping $v_x$, $v_y$, and $v_z$ positive and control constraints strictly satisfied.


\subsection{NMPC with GMRES Algorithm}

This section denotes how the NMPC method finds optimal control inputs together with the GMRES scheme as given in \cite{Ohtsuka_A_continuation/GMRES_method_for_fast_computation_of_nonlinear_receding_horizon_control}. The NMPC approach solves the control problem given in Eq.~\eqref{NMPC_formulation} as follows. First, we discretize all the formulation, constraints, and the cost function.

\begin{equation}\label{Real_time_Update_1}
    \boldsymbol{x}^*_{i+1}(t)=\boldsymbol{x}^*_{i}(t)+f(\boldsymbol{x}^*_{i}(t),\boldsymbol{u}^*_{i}(t))\Delta\tau
\end{equation}

\begin{equation}\label{Real_time_Update_2}
    \boldsymbol{x}^*_0(t)=\boldsymbol{x}(t)
\end{equation}

\begin{equation}\label{Real_time_Update_3}
    C(\boldsymbol{x}^*_{i}(t),\boldsymbol{u}^*_{i}(t))=0
\end{equation}

\begin{equation}
    J_\text{cost}=\psi(\boldsymbol{x}^*_N(t))+\sum^{N-1}_{i=0}L(\boldsymbol{x}^*_{i}(t),\boldsymbol{u}^*_{i}(t))\Delta\tau
\end{equation}
where $\Delta\tau=T/N$,
$f(\boldsymbol{x},\boldsymbol{u})$, $C(\boldsymbol{x},\boldsymbol{u})$, which is the equality constraint, and $L(\boldsymbol{x},\boldsymbol{u})$ are all defined as: 

\begin{equation}
    f(\boldsymbol{x},\boldsymbol{u})=
    \begin{bmatrix}
        \frac{1}{2}(q_4\omega_x -q_3\omega_y q_2\omega_z)\\
        \frac{1}{2}(q_3\omega_x +q_4\omega_y -q_1\omega_z)\\
        \frac{1}{2}(-q_2\omega_x +q_1\omega_y q_4\omega_z)\\
        \frac{1}{2}(-q_1\omega_x -q_2\omega_y -q_3\omega_z)\\
        \frac{1}{J_x}\{(J_y-J_z)\omega_y\omega_z+ m_yB_z(t)-m_zB_y(t)\}\\
        \frac{1}{J_y}\{(J_z-J_x)\omega_z\omega_x+m_zB_x(t)-m_xB_z(t)\}\\
        \frac{1}{J_z}\{(J_x-J_y)\omega_x\omega_y+m_xB_z(t)-m_yB_x(t)\}\\
    \end{bmatrix}
\end{equation}

\begin{equation}
    C(\boldsymbol{x},\boldsymbol{u})=
    \begin{bmatrix}
        m_x^2+v_x^2-u_\text{max}^2\\
        m_y^2+v_y^2-u_\text{max}^2\\
        m_z^2+v_z^2-u_\text{max}^2\\
    \end{bmatrix}
\end{equation}

\begin{equation}
    L(\boldsymbol{x},\boldsymbol{u})=
    \frac{1}{2}\{(\boldsymbol{x}(t+T)\boldsymbol{x}_{f})^{T}Q(\boldsymbol{x}(t+T)-\boldsymbol{x}_{f})+\boldsymbol{u}(t+T)^TR\boldsymbol{u}(t+T)-p_1v_1-p_2v_2-p_3v_3\}
\end{equation}
    
Let the initial state of the discretized problem be the current state vector as $\boldsymbol{x}^*_0(t)=\boldsymbol{x}(t)$, then a sequence of optimal control inputs $\{\boldsymbol{u}^*_i(t)\}^{N-1}_{i=0}$ is found at each sample time. At last, the control inputs that are actually given to the system is found as the first term of this sequence and is defined as $\boldsymbol{u}(t)=\boldsymbol{u}^*_0(t)$.

The solution to this discretized problem is found by introducing the
Hamiltonian, $H$, as
\begin{equation}
    H(\boldsymbol{x},\boldsymbol{\lambda},\boldsymbol{u},\boldsymbol{\mu})=L(\boldsymbol{x},\boldsymbol{u})+\boldsymbol{\lambda}^T f(\boldsymbol{x},\boldsymbol{u})+\boldsymbol{\mu}^T C(\boldsymbol{x},\boldsymbol{u})
\end{equation}
where $\boldsymbol{\lambda}$ is the vector of costate and $\boldsymbol{\mu}$ is the Lagrange multiplier associated with the equality constraint. The first-order necessary conditions for optimality~\cite{necessary_conditions_for_inputs} dictate that  $\{\boldsymbol{u}^*_i(t)\}^{N-1}_{i=0}$,
$\{\boldsymbol{\mu}^*_i(t)\}^{N-1}_{i=0}$,
$\{\boldsymbol{\lambda}^*_i(t)\}^{N-1}_{i=0}$, 
satisfy the following conditions:
\begin{equation}\label{Real_time_Update_4}
    H_{\boldsymbol{u}} (\boldsymbol{x}^*_{i}(t),\boldsymbol{\lambda}^*_{i+1}(t),\boldsymbol{u}^*_{i}(t),\boldsymbol{\mu}^*_{i}(t))=0
\end{equation}

\begin{equation}\label{Real_time_Update_5}
    \boldsymbol{\lambda}^*_i(t)=\boldsymbol{\lambda}^*_{i+1}(t)+H_{\boldsymbol{x}}^T(\boldsymbol{x}^*_{i}(t),\boldsymbol{\lambda}^*_{i+1}(t),\boldsymbol{u}^*_{i}(t),\boldsymbol{\mu}^*_{i}(t))\Delta\tau
\end{equation}

\begin{equation}\label{Real_time_Update_6}
    \boldsymbol{\lambda}^*_N(t)=\psi^T_{\boldsymbol{x}}(\boldsymbol{x}^*_N(t))
\end{equation}

The optimal problem is finally shown as a two-point boundary-value problem (TPBVP) for the discretized optimal control problem, where $\{\boldsymbol{u}^*_i(t)\}^{N-1}_{i=0}$ and $\{\boldsymbol{\mu}^*_i(t)\}^{N-1}_{i=0}$ satisfy Eqs.(\ref{Real_time_Update_1}--\ref{Real_time_Update_3}) and (\ref{Real_time_Update_4}--\ref{Real_time_Update_6}). 
To solve this TPBVP in a sufficiently small time, we apply GMRES method \cite{Ohtsuka_2} so as to update a new control input. 

\subsection{PWM}

As the NMPC controller described above creates continuous control inputs, a method that converts a smooth curve into a discrete signal helps to reduce actuator's burden. Therefore, we equip PWM method, which rounds off the smooth inputs to the nearest discrete values. In addition, to avoid unnecessary fluttering, we employed a algorithm suggested in \cite{Ohtsuka_hover}. This methodology uses a previous input value to determine the input at the current time step and prevents unrealistic fluctuation of control inputs. The algorithm below determines the optimal discrete value depending on the input at the previous sampling moment; the algorithm tends to keep the same control value as the preceding time step value. The algorithm sorts the smooth value into either ${u}_{\rm max}$, $\frac{2}{3}{u}_{\rm max}$, $\frac{1}{3}{u}_{\rm max}$, $0$, $-\frac{1}{3}{u}_{\rm max}$, $-\frac{2}{3}{u}_{\rm max}$, or $-{u}_{\rm max}$. Let $u_c$ be the NMPC continuous inputs, $u_d$ be discrete-valued inputs, $u_{pr}$ be the input at the previous sampling time, and $u_{\rm span}$ be $\frac{1}{3}u_{\rm max}$. Introducing a positive constant $\kappa$, we here show the condition branching algorithm of PWM as the following. \\

If ${u}_{pr}={u}_{\rm max}$:\\

\begin{equation}
    u_d=
    \begin{cases}
      u_{\rm max} & \text{if} \ \ u_c \geq \frac{2}{3}u_{\rm max}+(1-\kappa)u_{\rm span}/2 \\
      \frac{2}{3}u_{\rm max} & \text{if} \ \ \frac{1}{3}u_{\rm max}+(1+\kappa)u_{\rm span}/2 \leq u_c < \frac{2}{3}u_{\rm max}+(1-\kappa)u_{\rm span}/2 \\
      \frac{1}{3}u_{\rm max} & \text{if} \ \ 0+u_{\rm span}/2 \leq u_c < \frac{1}{3}u_{\rm max}+(1+\kappa)u_{\rm span}/2 \\
      0 & \text{if} \ \ -\frac{1}{3}u_{\rm max}+u_{\rm span}/2 \leq u_c < 0+u_{\rm span}/2 \\
      -\frac{1}{3}u_{\rm max} & \text{if} \ \ -\frac{2}{3}u_{\rm max}+u_{\rm span}/2 \leq u_c < -\frac{1}{3}u_{\rm max}+u_{\rm span}/2 \\
      -\frac{2}{3}u_{\rm max} & \text{if} \ \ -u_{\rm max}+u_{\rm span}/2 \leq u_c < -\frac{2}{3}u_{\rm max}+u_{\rm span}/2 \\
      -u_{\rm max} & \text{if} \ \ u_c < -u_{\rm max}+u_{\rm span}/2
    \end{cases}
\end{equation}

If ${u}_{pr}=\frac{2}{3}{u}_{\rm max}$:\\

\begin{equation}
    u_d=
    \begin{cases}
      u_{\rm max} & \text{if} \ \ u_c \geq \frac{2}{3}u_{\rm max}+(1+\kappa)u_{\rm span}/2 \\
      \frac{2}{3}u_{\rm max} & \text{if} \ \ \frac{1}{3}u_{\rm max}+(1+\kappa)u_{\rm span}/2 \leq u_c < \frac{2}{3}u_{\rm max}+(1+\kappa)u_{\rm span}/2 \\
      \frac{1}{3}u_{\rm max} & \text{if} \ \ 0+u_{\rm span}/2 \leq u_c < \frac{1}{3}u_{\rm max}+(1+\kappa)u_{\rm span}/2 \\
      0 & \text{if} \ \ -\frac{1}{3}u_{\rm max}+u_{\rm span}/2 \leq u_c < 0+u_{\rm span}/2 \\
      -\frac{1}{3}u_{\rm max} & \text{if} \ \ -\frac{2}{3}u_{\rm max}+u_{\rm span}/2 \leq u_c < -\frac{1}{3}u_{\rm max}+u_{\rm span}/2 \\
      -\frac{2}{3}u_{\rm max} & \text{if} \ \ -u_{\rm max}+u_{\rm span}/2 \leq u_c < -\frac{2}{3}u_{\rm max}+u_{\rm span}/2 \\
      -u_{\rm max} & \text{if} \ \ u_c < -u_{\rm max}+u_{\rm span}/2
    \end{cases}
\end{equation}

If ${u}_{pr}=\frac{1}{3}{u}_{\rm max}$:\\

\begin{equation}
    u_d=
    \begin{cases}
      u_{\rm max} & \text{if} \ \ u_c \geq \frac{2}{3}u_{\rm max}+u_{\rm span}/2 \\
      \frac{2}{3}u_{\rm max} & \text{if} \ \ \frac{1}{3}u_{\rm max}+(1+\kappa)u_{\rm span}/2 \leq u_c < \frac{2}{3}u_{\rm max}+u_{\rm span}/2 \\
      \frac{1}{3}u_{\rm max} & \text{if} \ \ 0+(1+\kappa)u_{\rm span}/2 \leq u_c < \frac{1}{3}u_{\rm max}+(1+\kappa)u_{\rm span}/2 \\
      0 & \text{if} \ \ -\frac{1}{3}u_{\rm max}+u_{\rm span}/2 \leq u_c < 0+(1+\kappa)u_{\rm span}/2 \\
      -\frac{1}{3}u_{\rm max} & \text{if} \ \ -\frac{2}{3}u_{\rm max}+u_{\rm span}/2 \leq u_c < -\frac{1}{3}u_{\rm max}+u_{\rm span}/2 \\
      -\frac{2}{3}u_{\rm max} & \text{if} \ \ -u_{\rm max}+u_{\rm span}/2 \leq u_c < -\frac{2}{3}u_{\rm max}+u_{\rm span}/2 \\
      -u_{\rm max} & \text{if} \ \ u_c < -u_{\rm max}+u_{\rm span}/2
    \end{cases}
\end{equation}

If ${u}_{pr}=0$:\\

\begin{equation}
    u_d=
    \begin{cases}
      u_{\rm max} & \text{if} \ \ u_c \geq \frac{2}{3}u_{\rm max}+u_{\rm span}/2 \\
      \frac{2}{3}u_{\rm max} & \text{if} \ \ \frac{1}{3}u_{\rm max}+u_{\rm span}/2 \leq u_c < \frac{2}{3}u_{\rm max}+u_{\rm span}/2 \\
      \frac{1}{3}u_{\rm max} & \text{if} \ \ 0+(1+\kappa)u_{\rm span}/2 \leq u_c < \frac{1}{3}u_{\rm max}+u_{\rm span}/2 \\
      0 & \text{if} \ \ -\frac{1}{3}u_{\rm max}+(1+\kappa)u_{\rm span}/2 \leq u_c < 0+(1+\kappa)u_{\rm span}/2 \\
      -\frac{1}{3}u_{\rm max} & \text{if} \ \ -\frac{2}{3}u_{\rm max}+u_{\rm span}/2 \leq u_c < -\frac{1}{3}u_{\rm max}+(1+\kappa)u_{\rm span}/2 \\
      -\frac{2}{3}u_{\rm max} & \text{if} \ \ -u_{\rm max}+u_{\rm span}/2 \leq u_c < -\frac{2}{3}u_{\rm max}+(1+\kappa)u_{\rm span}/2 \\
      -u_{\rm max} & \text{if} \ \ u_c < -u_{\rm max}+u_{\rm span}/2
    \end{cases}
\end{equation}

If ${u}_{pr}=-\frac{1}{3}{u}_{\rm max}$:\\

\begin{equation}
    u_d=
    \begin{cases}
      u_{\rm max} & \text{if} \ \ u_c \geq \frac{2}{3}u_{\rm max}+u_{\rm span}/2 \\
      \frac{2}{3}u_{\rm max} & \text{if} \ \ \frac{1}{3}u_{\rm max}+u_{\rm span}/2 \leq u_c < \frac{2}{3}u_{\rm max}+u_{\rm span}/2 \\
      \frac{1}{3}u_{\rm max} & \text{if} \ \ 0+u_{\rm span}/2 \leq u_c < \frac{1}{3}u_{\rm max}+u_{\rm span}/2 \\
      0 & \text{if} \ \ -\frac{1}{3}u_{\rm max}+(1+\kappa)u_{\rm span}/2 \leq u_c < 0+u_{\rm span}/2 \\
      -\frac{1}{3}u_{\rm max} & \text{if} \ \ -\frac{2}{3}u_{\rm max}+(1+\kappa)u_{\rm span}/2 \leq u_c < -\frac{1}{3}u_{\rm max}+(1+\kappa)u_{\rm span}/2 \\
      -\frac{2}{3}u_{\rm max} & \text{if} \ \ -u_{\rm max}+u_{\rm span}/2 \leq u_c < -\frac{2}{3}u_{\rm max}+(1+\kappa)u_{\rm span}/2 \\
      -u_{\rm max} & \text{if} \ \ u_c < -u_{\rm max}+u_{\rm span}/2
    \end{cases}
\end{equation}

If ${u}_{pr}=-\frac{2}{3}{u}_{\rm max}$:\\

\begin{equation}
    u_d=
    \begin{cases}
      u_{\rm max} & \text{if} \ \ u_c \geq \frac{2}{3}u_{\rm max}+u_{\rm span}/2 \\
      \frac{2}{3}u_{\rm max} & \text{if} \ \ \frac{1}{3}u_{\rm max}+u_{\rm span}/2 \leq u_c < \frac{2}{3}u_{\rm max}+u_{\rm span}/2 \\
      \frac{1}{3}u_{\rm max} & \text{if} \ \ 0+u_{\rm span}/2 \leq u_c < \frac{1}{3}u_{\rm max}+u_{\rm span}/2 \\
      0 & \text{if} \ \ -\frac{1}{3}u_{\rm max}+u_{\rm span}/2 \leq u_c < 0+u_{\rm span}/2 \\
      -\frac{1}{3}u_{\rm max} & \text{if} \ \ -\frac{2}{3}u_{\rm max}+(1+\kappa)u_{\rm span}/2 \leq u_c < -\frac{1}{3}u_{\rm max}+u_{\rm span}/2 \\
      -\frac{2}{3}u_{\rm max} & \text{if} \ \ -u_{\rm max}+(1+\kappa)u_{\rm span}/2 \leq u_c < -\frac{2}{3}u_{\rm max}+(1+\kappa)u_{\rm span}/2 \\
      -u_{\rm max} & \text{if} \ \ u_c < -u_{\rm max}+(1+\kappa)u_{\rm span}/2
    \end{cases}
\end{equation}

If ${u}_{pr}=-{u}_{\rm max}$:\\

\begin{equation}
    u_d=
    \begin{cases}
      u_{\rm max} & \text{if} \ \ u_c \geq \frac{2}{3}u_{\rm max}+u_{\rm span}/2 \\
      \frac{2}{3}u_{\rm max} & \text{if} \ \ \frac{1}{3}u_{\rm max}+u_{\rm span}/2 \leq u_c < \frac{2}{3}u_{\rm max}+u_{\rm span}/2 \\
      \frac{1}{3}u_{\rm max} & \text{if} \ \ 0+u_{\rm span}/2 \leq u_c < \frac{1}{3}u_{\rm max}+u_{\rm span}/2 \\
      0 & \text{if} \ \ -\frac{1}{3}u_{\rm max}+u_{\rm span}/2 \leq u_c < 0+u_{\rm span}/2 \\
      -\frac{1}{3}u_{\rm max} & \text{if} \ \ -\frac{2}{3}u_{\rm max}+(1+\kappa)u_{\rm span}/2 \leq u_c < -\frac{1}{3}u_{\rm max}+u_{\rm span}/2 \\
      -\frac{2}{3}u_{\rm max} & \text{if} \ \ -u_{\rm max}+(1-\kappa)u_{\rm span}/2 \leq u_c < -\frac{2}{3}u_{\rm max}+(1+\kappa)u_{\rm span}/2 \\
      -u_{\rm max} & \text{if} \ \ u_c < -u_{\rm max}+(1-\kappa)u_{\rm span}/2 
    \end{cases}
\end{equation}


\section{Simulation Results}

This section illustrates the simulation results of the detumbling and attitude control process of small satellites. We showcase the feasibility of the NMPC algorithm and PWM method on an asymmetric satellite. The simulations present the satellite's dynamics propagated on the sun-synchronous orbit, whose orbital elements are all given in Table~\ref{Six_elements}.

\subsection{Orbital Elements}

\begin{table}[H]
\caption{\label{Six_elements} Six elements of the sun synchronous orbit}
\centering
\begin{tabular}{ c c }
    \hline
    Semi-major axis & 6691.6 [km]\\
    \hline
    Eccentricity & 0.046440\\
    \hline
    Inclination & 96.700 [deg]\\
    \hline
    Right Ascension of Ascending Node &  100.90 [deg]\\
    \hline
    Argument of perigee &  119.70 [deg]\\
    \hline
    Mean anomaly &  240.49 [deg]\\
    \hline
\end{tabular}
\end{table}


\subsection{Earth's Magnetic Field Model}

\begin{figure}[H]
\centering
\includegraphics[width=1\textwidth]{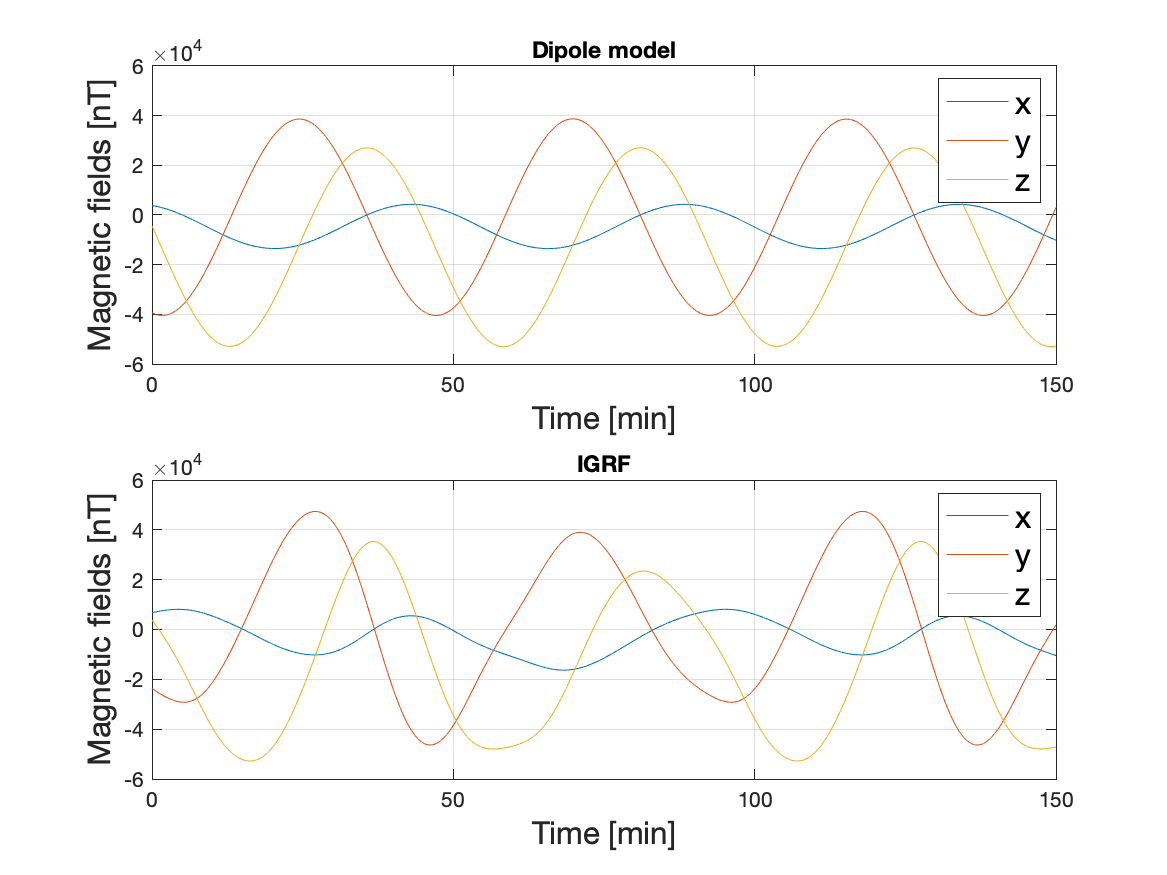}
\caption{Magnetic field in dipole and IGRF on the sun-synchronous orbit.}
\label{fig:magnetic_field}
\end{figure}

The simulation of satellites' rotational dynamics requires a model of Earth's magnetic field; therefore, we employ International Geomagnetic Reference Field (IGRF) \cite{IGRF} as an Earth's magnetic model in our simulation. On the other hand, applying IGRF to on-board calculation increases computational complexity and makes it less feasible. Hence, we equip a dipole magnetic model described in Eq.~\eqref{dipole_model} \cite{dipole_model} as an on-board magnetic model. The on-board NMPC controller refers to the dipole model and optimizes control inputs based on predicted future dynamics.

\begin{equation}\label{dipole_model}
\begin{array}{c}
    \begin{bmatrix}
    B_0x\\
    B_0y\\
    B_0z\\
    \end{bmatrix}
    =D_m
    \begin{bmatrix}
    \frac{3}{2}\sin{i}\sin{2\eta}\\
    -\frac{3}{2}\sin{i}\left(\cos{2\eta}-\frac{1}{3}\right)\\
    -\cos{i}\\
    \end{bmatrix}
\end{array}
\end{equation}
where $\eta = \theta+\omega_{e}$, $D_m=-\frac{M_e}{r^3}$, $M_e=8.1\times10^{25}$ [gauss $\cdot$ cm$^{3}$], $r$ is a distance between the satellite and the center of the Earth, $\theta$ is true anomaly, and $\omega_e$ is argument of perigee. Figure \ref{fig:magnetic_field} compares the two models on the sun-synchronous orbit. 

\subsection{PWM Detumbling}

This section demonstrates NMPC's capability of detumbling all the three-axis angular velocity. A small satellite's moment of inertia is given in the Table~\ref{moment_of_inertia}.

\begin{table}[H]
\caption{\label{moment_of_inertia} Moment of inertia of the Asymmetric satellite }
\centering
\begin{tabular}{ c c }
    \hline
    Moment of inertia & Value [kg $\cdot$ m$^2$]\\
    \hline
    $J_x$ & 0.020 \\ 
    \hline
    $J_y$ & 0.030 \\ 
    \hline
    $J_z$ & 0.040 \\ 
    \hline
\end{tabular}
\end{table}

Table~\ref{initial_condition_and_ref} gives the initial conditions and the reference states, where the state vector is $\boldsymbol{x}=[q_1, q_2, q_3, q_4, \omega_x, \omega_y, \omega_z]^{T}$.

\begin{table}[H] 
\caption{\label{initial_condition_and_ref} The initial condition and the reference states}
\centering
\begin{tabular}{c c}
    \hline
    Initial condition, $\boldsymbol{x}_0$ & [0, 0, 0, 1, 3 [deg], 3 [deg], 3 [deg]]$^{T}$ \\
    \hline
    Reference state, $\boldsymbol{x}_f$ & [0, 0, 0, 1, 0 [deg], 0 [deg], 0 [deg]]$^{T}$ \\ 
    \hline
\end{tabular}
\end{table}

The NMPC and PWM properties are all given in Table~\ref{table_NMPC_properties_asymmetric}. 

\begin{table}[H] 
\caption{\label{table_NMPC_properties_asymmetric} NMPC and PWM properties}
\centering
\begin{tabular}{c c}
    \hline
    maximum control input, $u_{\rm max}$ & 0.10 [A m$^2$] \\
    \hline
    $T_s$ & 10 [sec] \\
    \hline
    \textit{$Q$} & diag([0, 0, 0, 0, 100, 100, 250])  \\ 
    \hline
    \textit{$Q_{\rm t}$} &  diag([0, 0, 0, 0, 100, 100, 250]) \\ 
    \hline
    \textit{$R$} & diag([0, 0, 0, 10$^{-8}$, 10$^{-8}$, 10$^{-8}$, 10$^{-8}$, 10$^{-8}$, 10$^{-8}$]) \\ 
    \hline
    \textit{$p$} & diag([$10^{-1}$, $10^{-1}$, $10^{-1}$]) \\  
    \hline
    $N$ & $10$ \\
    \hline
    $\Delta \tau$ & 1.0 [sec] \\
    \hline
    $\kappa$ & 0.30 \\
    \hline
\end{tabular}
\end{table}
where $T_s$ is prediction horizon, $Q$, $Q_{\rm t}$, $R$, and $p$ are weight matrices, $N$ is discretized step number on prediction horizon, $\Delta \tau=T_s/N$, and $\kappa$ is the PWM method's constant, which prevents unnecessary fluctuation of inputs. 

\begin{figure}[H]
\centering
\includegraphics[width=1\textwidth]{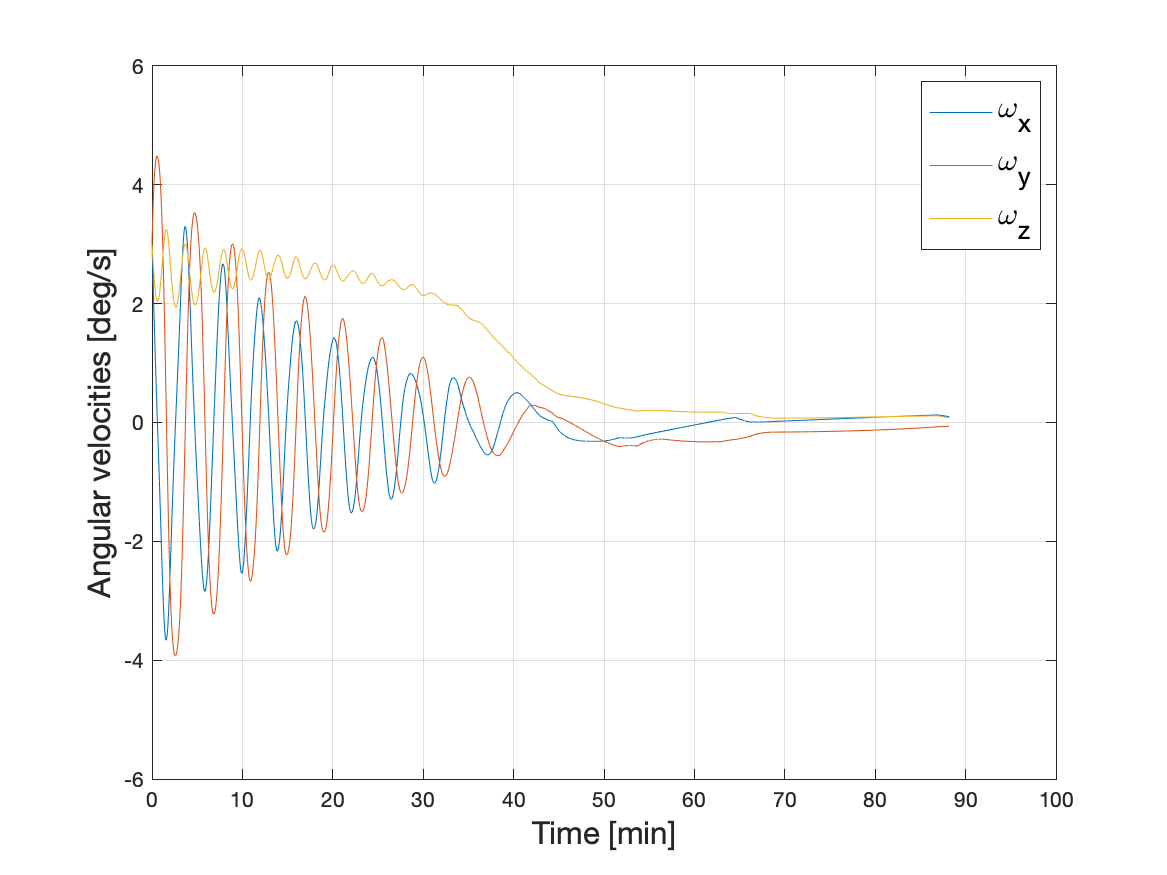}
\caption{Time history of angular velocities on NMPC on sun-synchronous orbit.}
\label{fig:PWM_detumbling_NMPC}
\end{figure}

Figure.~\ref{fig:PWM_detumbling_NMPC} illustrates the NMPC controller achieves stabilization within 100 minutes. The simulation is terminated when all the components' magnitude of angular rate is less than 0.10 [deg/s]. Intriguingly, the controller first makes the angular velocities larger. In the end, however, all the components are stabilized.

\begin{figure}[H]
\centering
\includegraphics[width=1\textwidth]{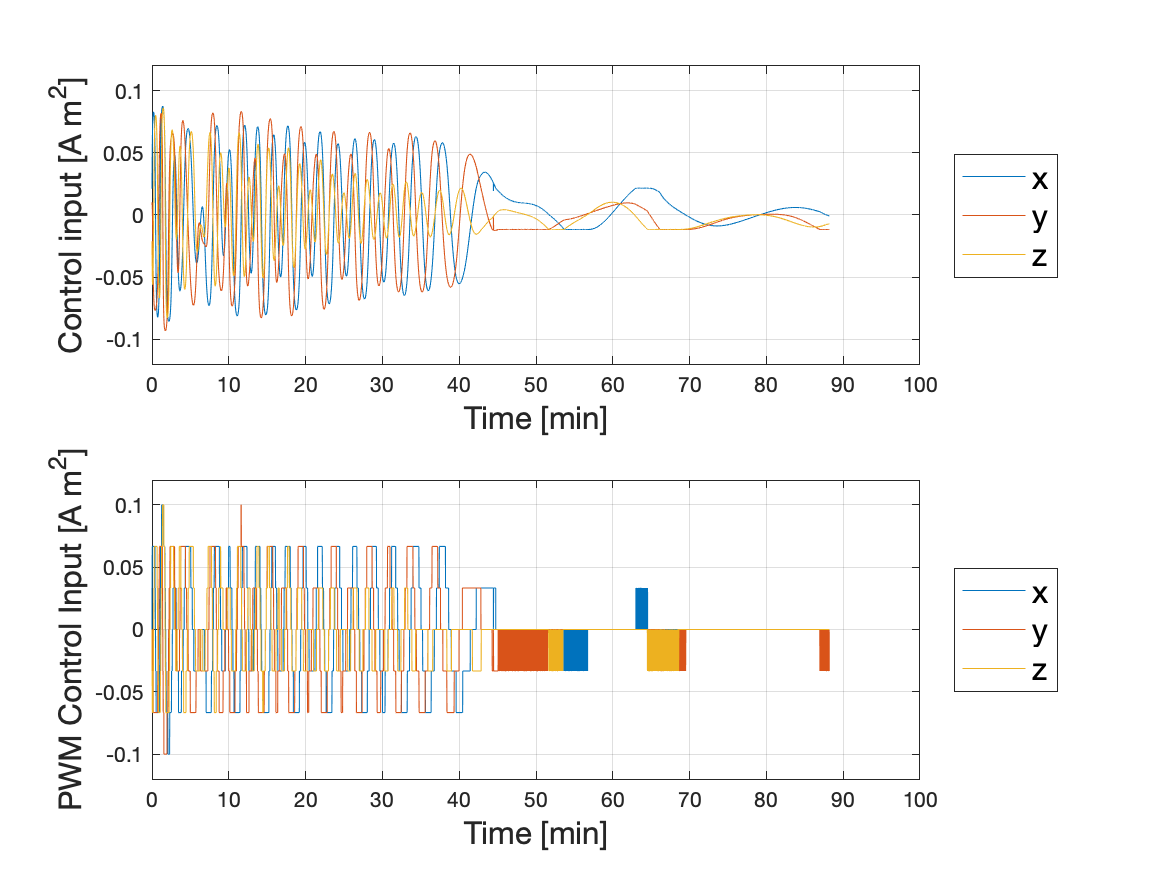}
\caption{Time history of continuous and discrete control inputs}
\label{fig:PWM_detumbling_NMPC_input_comparison}
\end{figure}

The NMPC's continuous inputs are transformed by the PWM method into discrete inputs, as shown in Fig~\ref{fig:PWM_detumbling_NMPC_input_comparison}. The first graph shows continuous control inputs that the NMPC controller finds, and the second chart presents discrete inputs converted by the PWM method.   

\begin{figure}[H]
\centering
\includegraphics[width=1\textwidth]{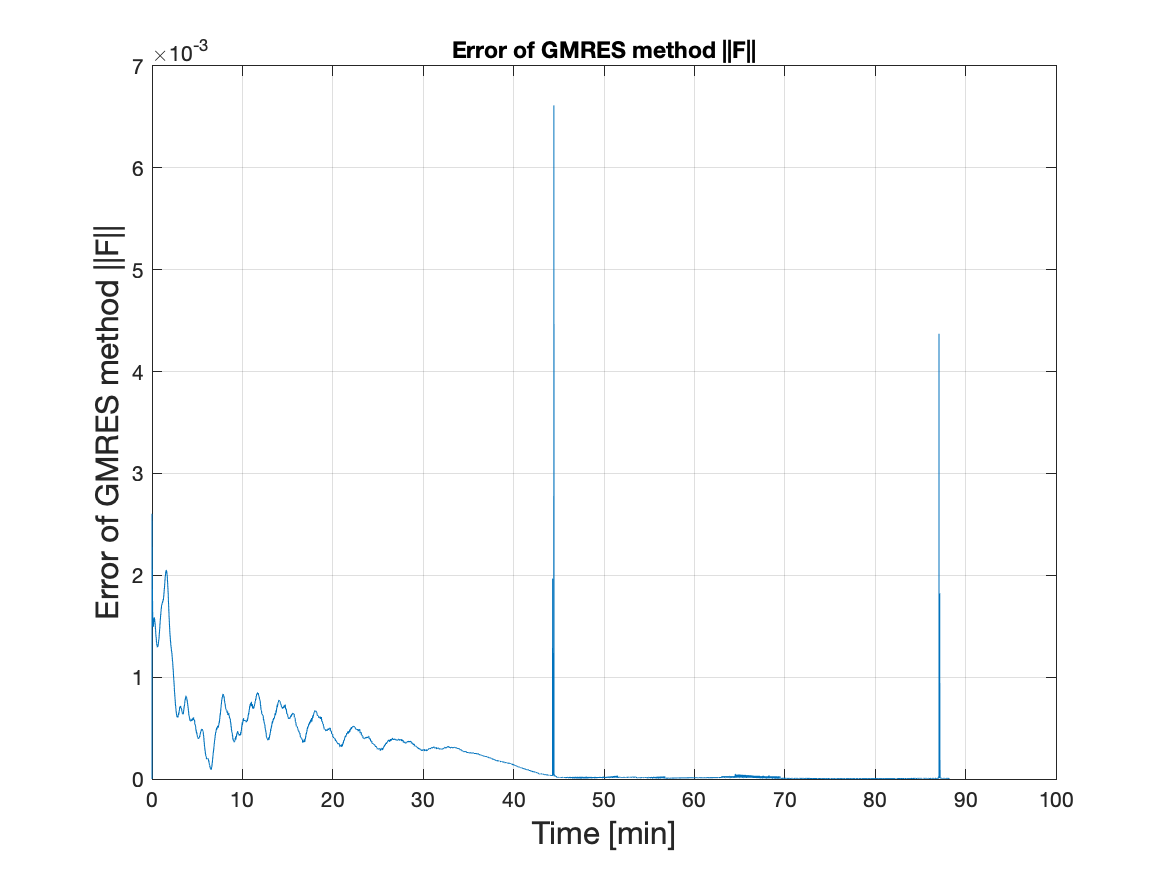}
\caption{The GMRES method error ||F|| }
\label{fig:PWM_detumbling_NMPC_error}
\end{figure}

Figure~\ref{fig:PWM_detumbling_NMPC_error} presents the error resulting from the GMRES method. The GMRES approaches find the optimal inputs within a constant iteration and period, but at the same time, it allows some value of calculation error. When the error, $||F|||$, becomes large, the controller fails to find the optimal solution, and hence, it is crucial to keep it small. In our case, the error is less than $7.0 \times 10^{-3}$, which is small enough for the NMPC methodology to find accurate control inputs. Note that there are periodic spikes around 44 [min] and 88 [min], and this might result from the discrepancy of the two magnetic models, as shown in \ref{fig:magnetic_field}.


\subsection{PWM Attitude Control}

This section shows attitude control simulation results. Table~\ref{initial_condition_and_ref_att} gives the initial conditions and the reference states, where the state vector is $\boldsymbol{x}=[q_1, q_2, q_3, q_4, \omega_x, \omega_y, \omega_z]^{T}$. Note that the initial angular rate is 0, whereas the initial quaternion differs from the reference state.  

\begin{table}[H] 
\caption{\label{initial_condition_and_ref_att} The initial condition and the reference states}
\centering
\begin{tabular}{c c}
    \hline
    Initial condition, $\boldsymbol{x}_0$ & [1, 0, 0, 0, 0 [deg], 0 [deg], 0 [deg]]$^{T}$ \\
    \hline
    Reference state, $\boldsymbol{x}_f$ & [0, 0, 0, 1, 0 [deg], 0 [deg], 0 [deg]]$^{T}$ \\ 
    \hline
\end{tabular}
\end{table}

The NMPC and PWM method properties are all given in Table~\ref{table_NMPC_properties_asymmetric_att}.

\begin{table}[H] 
\caption{\label{table_NMPC_properties_asymmetric_att} NMPC and PWM properties}
\centering
\begin{tabular}{c c}
    \hline
    maximum control input, $u_{\rm max}$ & 0.10 [A m$^2$] \\
    \hline
    $T_s$ & 5 [sec] \\
    \hline
    \textit{$Q$} & ${\rm diag}([20, 20, 20, 20, 2 \times 10^{4}, 2 \times 10^{4}, 2 \times 10^{4}])$ \\ 
    \hline
    \textit{$Q_{\rm t}$} & ${\rm diag}([100, 100, 100, 100, 2 \times 10^{4}, 2 \times 10^{4}, 2 \times 10^{4}])$ \\
    \hline
    \textit{$R$} & diag([0, 0, 0, $10^{-8}$, $10^{-8}$, $10^{-8}$, $10^{-8}$, $10^{-8}$, $10^{-8}$]) \\ 
    \hline
    \textit{$p$} & diag([$10^{-1}$, $10^{-1}$, $10^{-1}$]) \\  
    \hline
    $N$ & $20$ \\
    \hline
    $\Delta \tau$ & 0.25 [sec] \\
    \hline
    $\kappa$ & 0.30 \\
    \hline
\end{tabular}
\end{table}
where $T_s$ is prediction horizon, $Q$, $Q_{\rm t}$, $R$, and $p$ are weight matrices, $N$ is discretized step number on prediction horizon, $\Delta \tau=T_s/N$, and $\kappa$ is the PWM constant, which prevents unnecessary fluctuation of inputs. Note that $\Delta \tau$ is much smaller than that of the detumbling's, which makes the optimization more accurate.

\begin{figure}[H]
\centering
\includegraphics[width=1\textwidth]{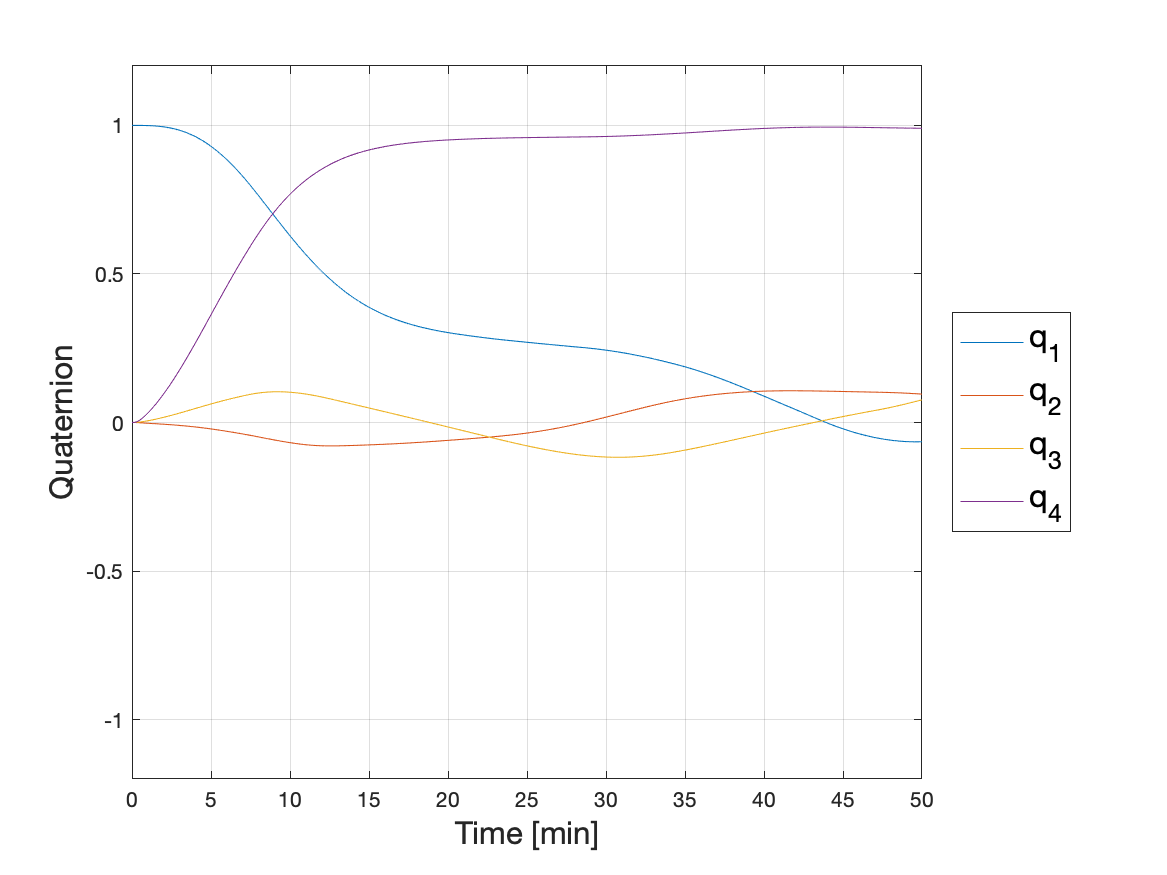}
\caption{Time history of quaternion on sun-synchronous orbit.}
\label{fig:attitude_control}
\end{figure}

As can be seen, the NMPC controller achieves attitude maneuver within 50 minutes, if not perfectly. At 50 [min], $q_4$ is larger than 0.99, and all the other components are less than 0.10. Table~\ref{table_att_at_the_end_quaternions} gives all the quaternion components at 50 [min]. Note that there is a good chance that by adjusting NMPC parameters, this maneuver can be accomplished for much smaller amount of time.

\begin{table}[H] 
\caption{\label{table_att_at_the_end_quaternions} Quaternions at 50 [min]}
\centering
\begin{tabular}{c c c c}
    \hline
    $q_1$ & $q_2$ & $q_3$ & $q_4$ \\
    \hline
    -0.0647 & 0.0956 & 0.0763 & 0.9904 \\
    \hline
\end{tabular}
\end{table}

\begin{figure}[H]
\centering
\includegraphics[width=1\textwidth]{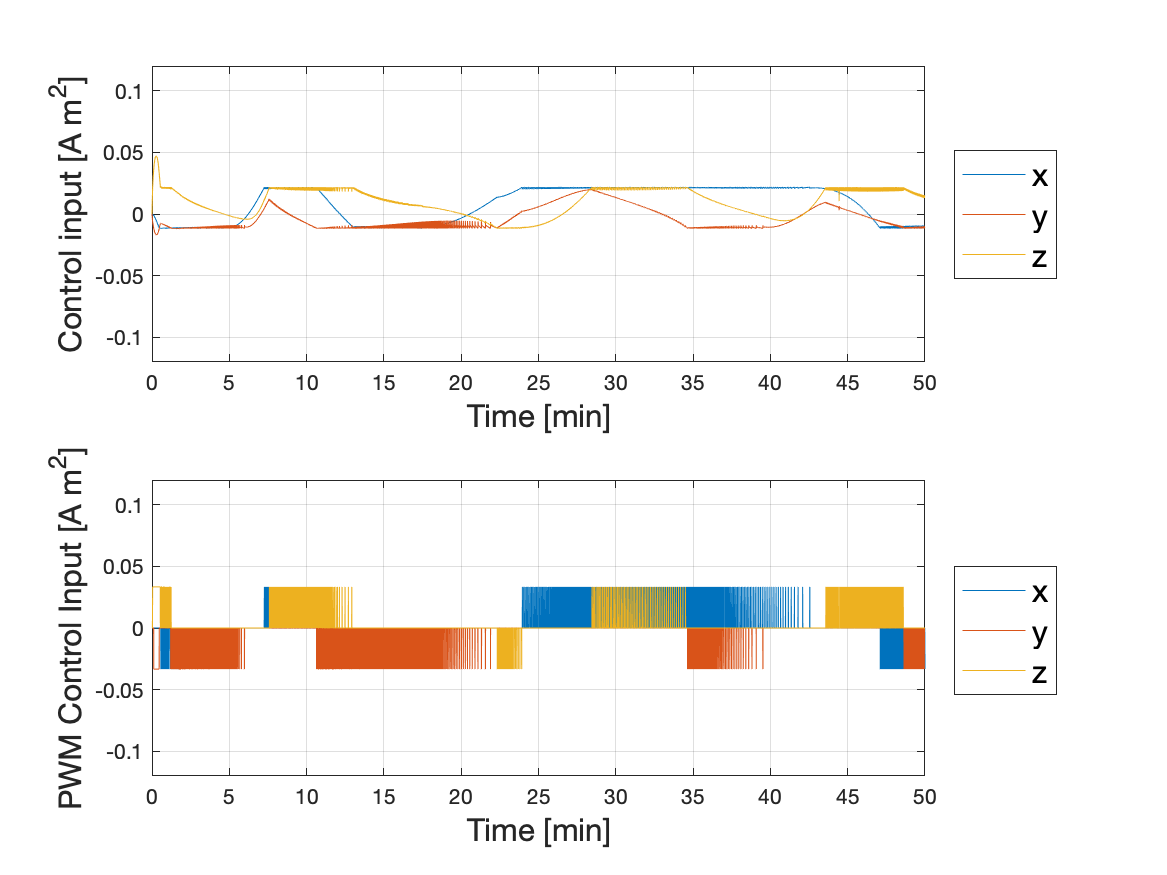}
\caption{Time history of continuous and discrete control inputs}
\label{fig:PWM_attitude_control_NMPC_input_comparison}
\end{figure}

The chart presents the difference between the continuous control inputs and the discrete inputs that was converted by the PWM scheme.   

\begin{figure}[H]
\centering
\includegraphics[width=1\textwidth]{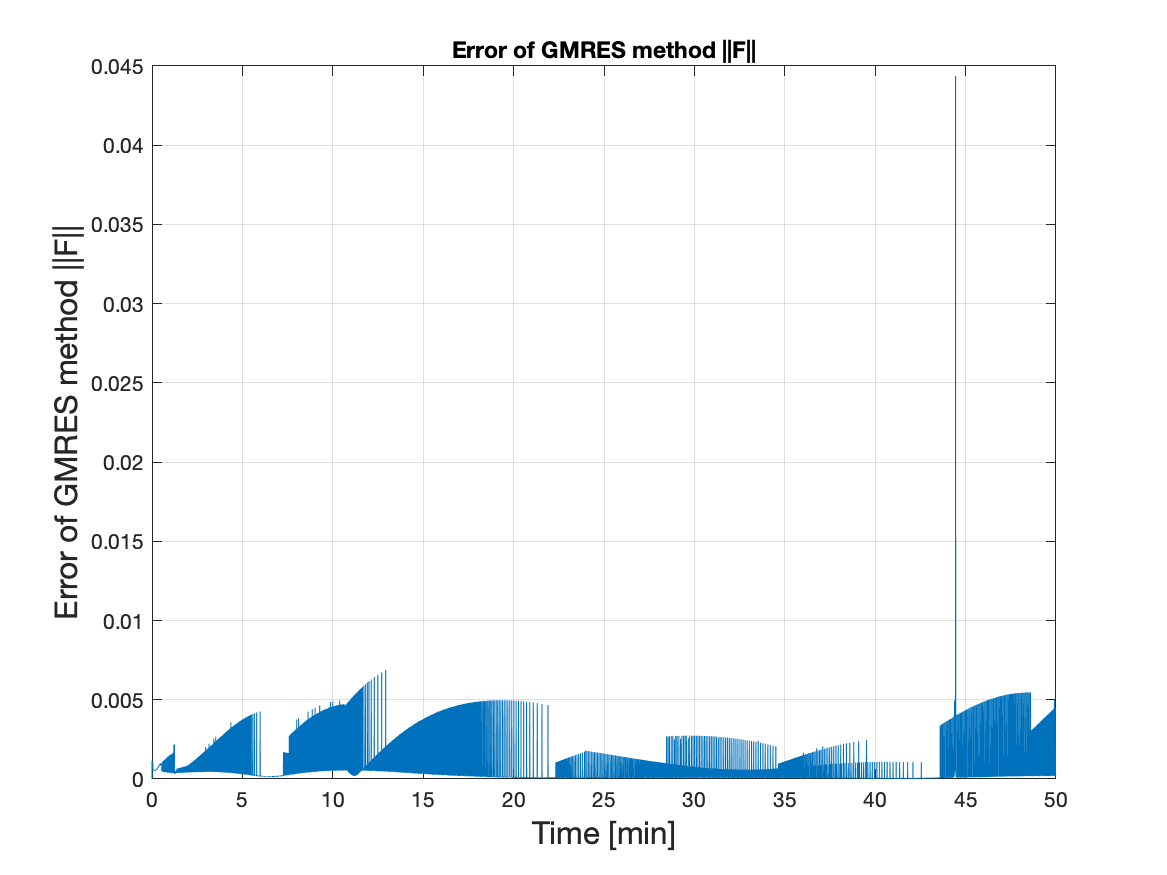}
\caption{The GMRES method error ||F|| }
\label{fig:PWM_attitude_control_NMPC_error}
\end{figure}

As with the case with detumbling given in Fig.~\ref{fig:PWM_detumbling_NMPC_error}, Fig.~\ref{fig:PWM_attitude_control_NMPC_error} indicates periodic spikes around 44 [min].

\subsection{Attitude Control with Continuous Input}

Figure.~\ref{fig:attitude_control} shows PWM's difficulty in achieving attitude control. Although quaternion components are close to the reference state, they do not converge within the given period. This section investigates whether the seen obstacle is inherent to the magnetic control system or caused by the PWM method’s discretization. We applied continuous control inputs to the system, and the results, whose NMPC properties are the same as given in Table~\ref{table_NMPC_properties_asymmetric_att}, are as follows. 

\begin{figure}[H]
\centering
\includegraphics[width=1\textwidth]{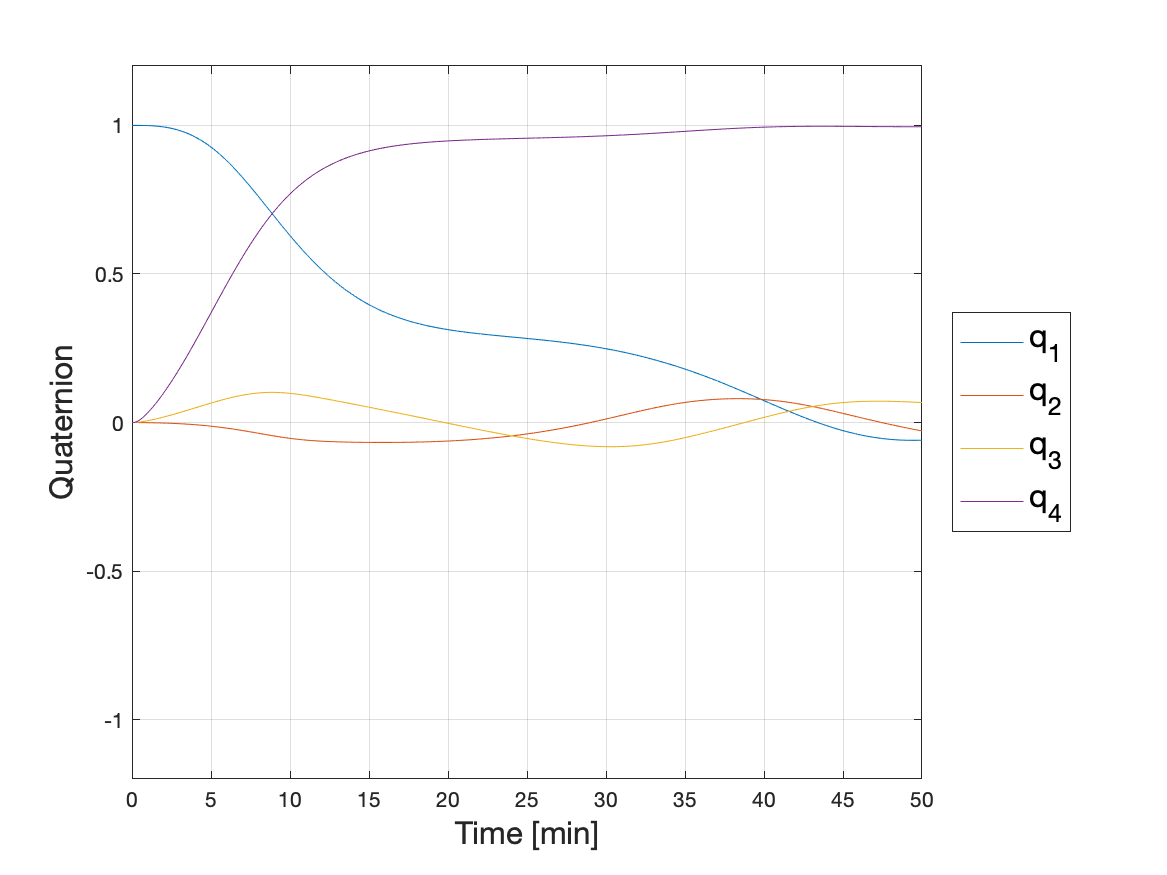}
\caption{Time history of quaternion on sun-synchronous orbit.}
\label{fig:continuous_attitude_control}
\end{figure}
The NMPC controller achieves attitude maneuver within 50 minutes; however, the same with the PWM-ed control, even the continuous input cannot make all the quaternion converge. It indicates that the discrete control does not trigger the difficulty in attitude maneuver, but instead, it is inherent to this magnetic-actuated satellite control system. Table~\ref{table_con_att_at_the_end_quaternions} gives all the quaternion components at 50 [min].

\begin{table}[H] 
\caption{\label{table_con_att_at_the_end_quaternions} Quaternions at 50 [min]}
\centering
\begin{tabular}{c c c c}
    \hline
    $q_1$ & $q_2$ & $q_3$ & $q_4$ \\
    \hline
    -0.0594 & -0.0284 & 0.0671 & 0.9956 \\
    \hline
\end{tabular}
\end{table}

As Table~\ref{table_con_att_at_the_end_quaternions} indicates, the continuous inputs converge the quaterion components better than the PWM-ed inputs; nonetheless, the difference is not significant. 

\begin{figure}[H]
\centering
\includegraphics[width=1\textwidth]{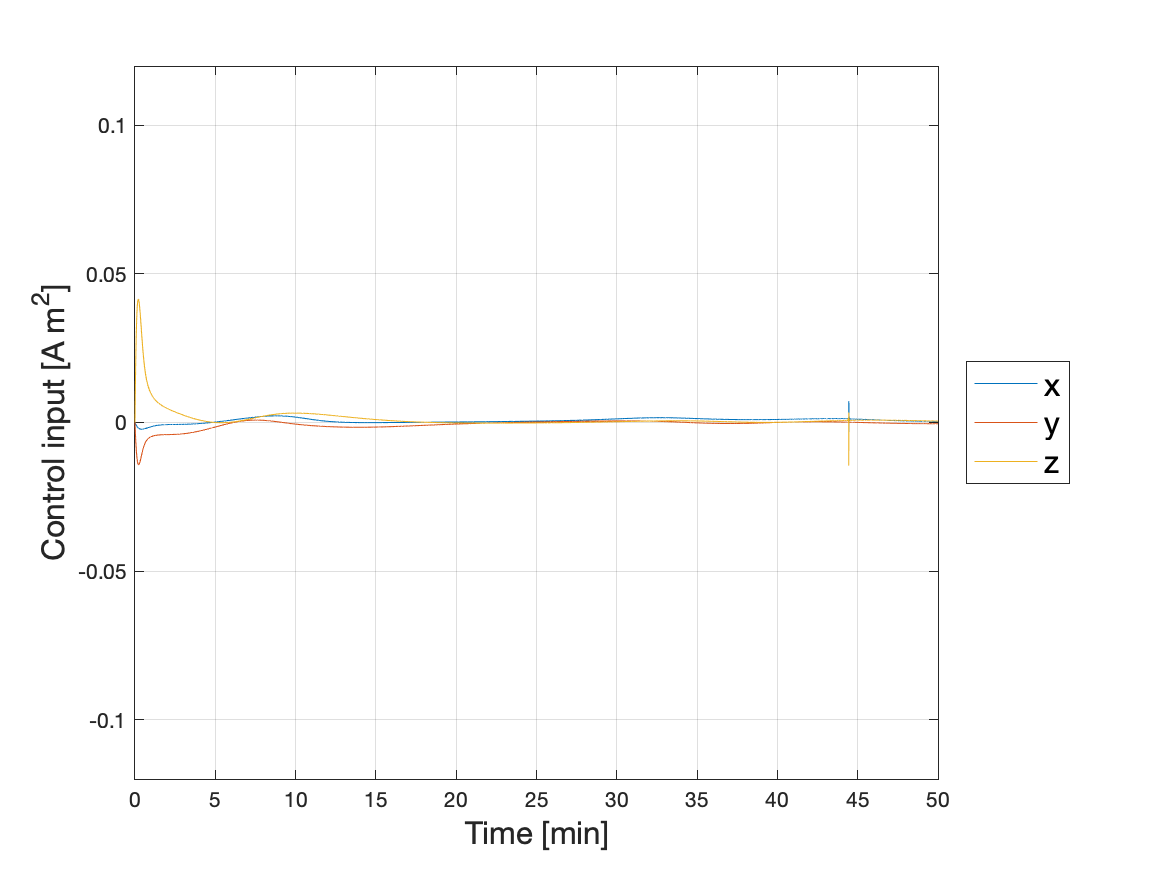}
\caption{Time history of continuous and discrete control inputs}
\label{fig:Continuous_attitude_control_NMPC_input_comparison}
\end{figure}

\begin{figure}[H]
\centering
\includegraphics[width=1\textwidth]{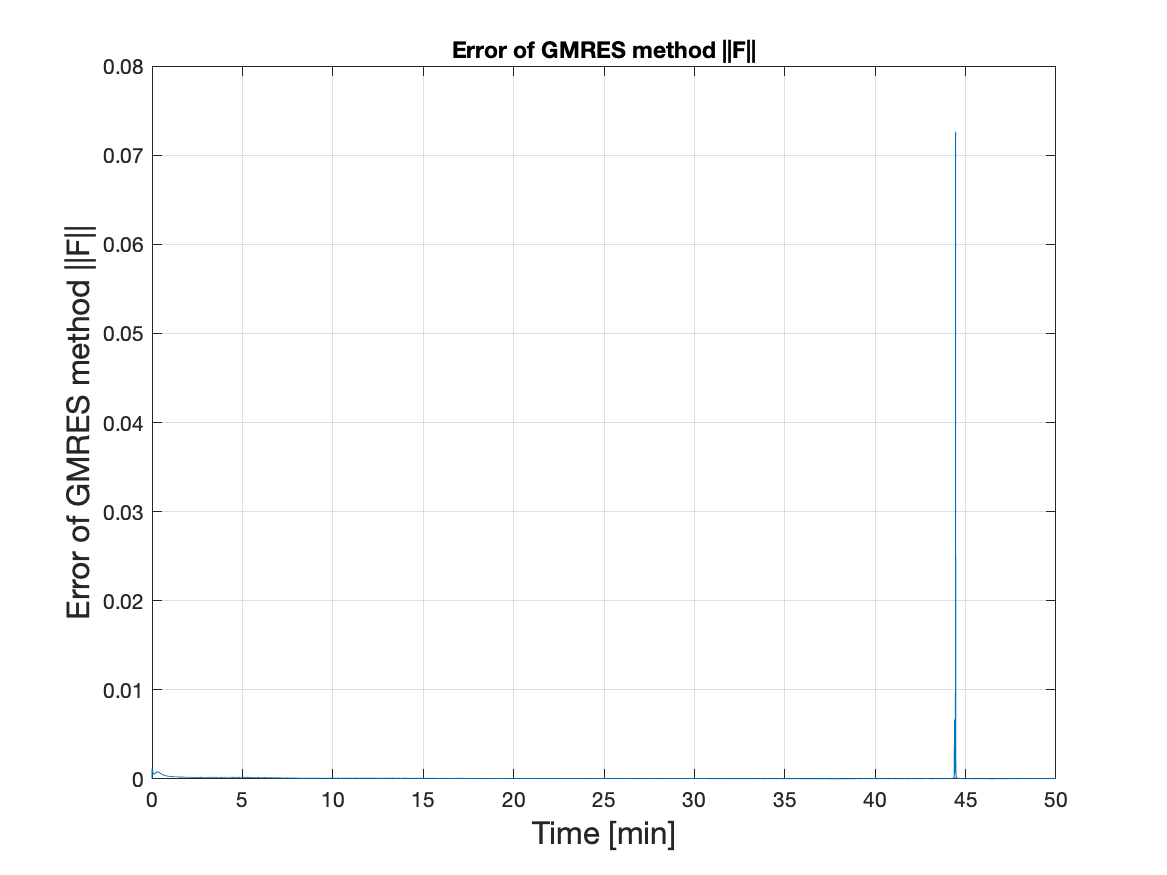}
\caption{The GMRES method error ||F|| }
\label{fig:Continuous_attitude_control_NMPC_error}
\end{figure}

Figure~\ref{fig:Continuous_attitude_control_NMPC_input_comparison} gives the continuous control inputs that the NMPC controller finds. The spike at 44 [min] corresponds to the sudden rise of the error shown in Fig.~\ref{fig:Continuous_attitude_control_NMPC_error}.

As discussed before, GMRES method's error is seen at 44 [min], and further research on this spike is needed for practical use of the NMPC, GMRES, and PWM on satellites. 

\section{Conclusions}

This paper has shown the PWM method’s applicability that converts smooth inputs found by the NMPC controller into discrete ones. The feedback scheme detumbles the small satellite and control its attitude with a three-axis magnetorquer. The PWM discretization reduces the complexity of the control actuator, which leads to a longer operation, the robustness of the system, and widens the range of application of actuators. Further research on the analysis of stability and simulations with different NMPC parameters, orbits, and spacecraft models would help investigate the proposed method. 

\bibliography{bib.bib}

\end{document}